\documentclass[sigconf]{acmart}

\usepackage{float}
\usepackage{multirow}

\AtBeginDocument{%
  \providecommand\BibTeX{{%
    \normalfont B\kern-0.5em{\scshape i\kern-0.25em b}\kern-0.8em\TeX}}}

\copyrightyear{2022}
\acmYear{2022}
\setcopyright{licensedothergov}\acmConference[ICSE-NIER'22]{New Ideas and Emerging Results }{May 21--29, 2022}{Pittsburgh, PA, USA}
\acmBooktitle{New Ideas and Emerging Results (ICSE-NIER'22), May 21--29, 2022, Pittsburgh, PA, USA}
\acmPrice{15.00}
\acmDOI{10.1145/3510455.3512775}
\acmISBN{978-1-4503-9224-2/22/05}

\begin{document}

\copyrightyear{2022}

\title{The best defense is a good defense: adapting negotiation methods for tackling pressure over software project estimates}

\author{Patricia G. F. Matsubara}
\email{patriciagfm@icomp.ufam.edu.br}
\affiliation{%
  \institution{Federal University of Amazonas (UFAM) \& Federal University of Mato Grosso do Sul (UFMS)}
  \city{Manaus-AM \& Campo Grande-MS}
  \country{Brazil}
}

\author{Igor Steinmacher}
\email{igorfs@utfpr.edu.br}
\affiliation{%
  \institution{Federal University of Technology - Paran{\'a} (UTFPR)}
  \city{Campo Mour\~{a}o-PR}
  \country{Brazil}
}

\author{Bruno Gadelha}
\email{bruno@icomp.ufam.edu.br}
\affiliation{%
  \institution{Federal University of Amazonas (UFAM)}
  \city{Manaus-AM}
  \country{Brazil}
}

\author{Tayana Conte}
\email{tayana@icomp.ufam.edu.br}
\affiliation{%
  \institution{Federal University of Amazonas (UFAM)}
  \city{Manaus-AM}
  \country{Brazil}
}

\renewcommand{\shortauthors}{Matsubara, et al.}

\begin{abstract}
Software estimation is critical for a software project’s success and a challenging activity. We argue that estimation problems are not restricted to the generation of estimates but also their use for commitment establishment: project stakeholders pressure estimators to change their estimates or to accept unrealistic commitments to attain business goals. In this study, we employed a Design Science Research (DSR) methodology to design an artifact based on negotiation methods, to empower software estimators in defending their estimates and searching for alternatives to unrealistic commitments when facing pressure. The artifact is a concrete step towards disseminating the soft skill of negotiation among practitioners. We present the preliminary results from a focus group that showed that practitioners from the software industry could use the artifact in a concrete scenario when estimating and establishing commitments about a software project. Our future steps include improving the artifact with the suggestions from focus group participants and evaluating it empirically in real software projects in the industry.
\end{abstract}



\maketitle

\section{Introduction}
Estimating software projects is a critical activity in software engineering. Software practitioners (here referred to as estimators) produce estimates about at least one parameter: size, effort, cost, or duration  \cite{mcconnell_software_2006}. Extensive research investigated the factors affecting estimators, and the generation of software estimates, including reasons for errors and strategies to improve accuracy \cite{basten_systematic_2014, halkjelsvik_time_2018}. 

We argue that the generation of estimates is not the only problem: the use of software estimates for commitments establishment in software projects also plays a crucial role in accuracy. Our argument builds on the fact that other project stakeholders, like upper management and clients (referred to as receptors), receive these estimates. Moreover, estimators and receptors must agree on commitments: promises to deliver a set of features on a deadline, at a specific level of quality \cite{mcconnell_what_2006}. Ideally, the technical estimates from estimators are the foundations to define such commitments. Nevertheless, ultimately, people develop and maintain software to satisfy desirable business outcomes---or targets \cite{mcconnell_what_2006}. Usually, receptors want to hit these targets. Previous research has shown that sometimes estimators are pressured to change their estimates \cite{magazinius_investigating_2012} to meet targets when the software estimates collide with them \cite{dagnino_estimating_2013}, making estimates acceptable to receptors, or “defensible” \cite{matsubara_buying_2021}. In addition, time pressure---the perception that time is scarce considering the tasks' demands---is standard in the software industry, and commercial pressure is behind many of its causes \cite{kuutila_what_2021}. Time pressure can increase efficiency up to a certain point and on the cost of an efficiency-quality trade-off \cite{kuutila_what_2021}. This reverberates in surveys in the software industry, where tight deadlines were the most cited and most likely cause for technical debt \cite{ramac_common_2020, rios_practitioners_2020}. In summary, when estimators fail to defend their estimates when facing pressure, consequences go beyond inaccurate estimates: software quality also suffers. 

Unfortunately, software practitioners do not possess the skills needed to truly defend their estimates from pressure \cite{mcconnell_politics_2006}. In this context, we aim to answer the following research question: \textbf{How to empower software practitioners to resist pressure and defend their software estimates during commitment establishment?} In the next section, we argue for the adaptation of negotiation methods to the estimation context to answer this research question. This is an innovative concrete step to enable software practitioners to develop the relevant soft skill of negotiation. Responding to the call to promote soft skills in Software Engineering \cite{capretz_call_2018}, we contribute to close the gaps regarding what the software industry needs from practitioners \cite{garousi_aligning_2019}, and to address essential soft skills that complement technical skills \cite{capretz_soft_2017}.  

\section{Background}
\label{sec:back}

According to \citet{schneider_definition_2017}, negotiation encompasses any situation that involves two or more parties that (i) consider the situation a negotiation and prepare accordingly; (ii) try to gain something with the interaction or to improve their situation; (iii) need each other to get their purpose; (iv) communicate back-and-forth, questioning each other, sharing information, and making offers or creating options; and (v) work towards an agreement. We argue that the interaction of the estimation process and commitment establishment in software projects presents many of the elements of this definition. For instance, when estimating, people prepare to negotiate, with some stakeholders increasing the estimates while others do the opposite \cite{magazinius_investigating_2012}. 

Next, we summarize the three negotiation methods that form our theoretical foundations. We start with principled negotiation \cite{fisher_getting_2011}: a groundbreaking method \cite{ogilvie_what_2008}. It shifted the approach from bargaining to a more flexible one \cite{hak_principled_2018}, revolutionized the teaching of negotiation in many different fields \cite{menkel-meadow_why_2006}, and impacted how practitioners think about negotiation \cite{tsay_decision-making_2009}. The other two methods build upon principled negotiation, providing additional guidelines when people insist on uncooperative behaviors \cite{ury_getting_2007}, or make unacceptable demands or requests \cite{ury_power_2012}. Collectively, the principles and steps of such methods have impacted a wide variety of domains: from health care (e.g., pediatric operating rooms \cite{sinskey_applying_2019}), to personal improvement (e.g., recommendations for raising interpersonal assertiveness \cite{ames_interpersonal_2017}), to military (e.g., practical guide for negotiating in the military \cite{eisen_jr_practical_2019}), and others.

\textbf{Principled negotiation:} Principled negotiation has four principles \cite{fisher_getting_2011}. First, \textbf{separate the people from the problem}: account for differences in thinking, emerging feelings, and misinterpretations during negotiation. Second, \textbf{focus on interests, not positions}. Focusing on positions --- like predefined deadlines --- can create an impasse. The other side’s desires and concerns --- their interests --- helps to uncover other positions that satisfy them. Third, \textbf{invent options for mutual gains}. The options should explore differences in parties’ interests. A convincing rationale and the consequence for each option also paves the way towards an agreement. Fourth, \textbf{insist on using objective criteria}: if interests continue to conflict, objective criteria support reaching an agreement based on principles instead of pressure.

\textbf{Breakthrough strategy:} Ury \cite{ury_getting_2007} proposed the breakthrough strategy for situations where people were especially uncooperative. The first step is to \textbf{go to the balcony} to suspend one's natural reactions (like giving in) when the other side refuses to reach an agreement. Next, one needs to \textbf{step to their side} by hearing them instead of arguing in the face of their disagreement and by expressing one's viewpoints without provocations. The third step is to \textbf{reframe}: redirect their attention from the positions to interests, creative options, and fair standards. The fourth step is to \textbf{build them a golden bridge} to walk out from their position to the agreement one wants: ask for constructive criticism of one's proposals or offer them choices to select from. If they keep rejecting, the fifth step is \textbf{to use power to educate}: let the other side know the consequences of no agreement.

\textbf{Positive no:} Sometimes, one must say no to a request because it is unacceptable. Poor reactions to such situations are to: (i) accommodate, saying yes when one feels like saying no; (ii) attack, saying no poorly and damaging the relationship; or (iii) avoid, saying nothing at all. Ury \cite{ury_power_2012} proposed the positive no method as an alternative. Its first stage is to \textit{prepare}, starting with to \textbf{uncover your Yes}: the reasons why one wants to say No. Next, one needs to \textbf{empower your No}, by devising a Plan B: a course of action one will take independently from the other side if they refuse to reach an agreement. The last step to \textbf{respect your way to Yes}, by trying to understand the other side utterly. The second stage is to \textit{deliver} one's No. One needs to \textbf{express your Yes}, clarifying one's motives for saying No. The next step is to \textbf{assert your No}. The final step is to \textbf{propose a Yes}, such as a third option to reconcile interests. The third stage is to \textit{follow through}. If the other side rejects the No, the next step is to \textbf{stay true to your Yes}: acknowledging their point of view without making concessions. Next, one needs \textbf{to underscore your No}. Let reality be their teacher about what will happen if they do not respect one's interests. If necessary, deploy one's Plan B respectfully. Lastly, one needs to \textbf{negotiate to Yes}, looking for their interests that might still be unmet by one's proposals.

\textbf{Related work:} McConnell \cite{mcconnell_politics_2006} argued in favor of using principled-negotiation \cite{fisher_getting_2011} in the discussions of estimates, targets, commitments, and plans --- also, to defend unpopular schedules \cite{mcconnell_how_1996}. Although the author provided experienced advice in the form of tips, he does not present a structured method completely adapted for the software estimation context. Also, the author did not empirically evaluate the tips. Moreover, other supporting negotiation methods emerged in the last years, such as the breakthrough strategy \cite{ury_getting_2007} and the positive no-method \cite{ury_power_2012}. In another study, Ochoa, Pino, and Poblete \cite{ochoa_estimating_2009} proposed a method for estimating software cost and duration, claiming to include negotiation as part of it. However, they only focused on internal team consensus, with no guidance about how to deal with external pressures that can lead the team to change their estimates. There is also no discussion about the negotiation theoretical foundations of the technique. To address the gaps from previous research, we proposed the creation and empirical evaluation of an artifact \cite{fernandes_matsubara_dealing_2019} based on the principled negotiation \cite{fisher_getting_2011}, the breakthrough strategy \cite{ury_getting_2007}, and the positive no method \cite{ury_power_2012}. Next, we describe our artifact and our preliminary results.  

\section{Defense Lenses for the Estimation Context}
\label{sec:lenses}

\begin{figure}[htb]
  \centering
  \includegraphics[width=\linewidth]{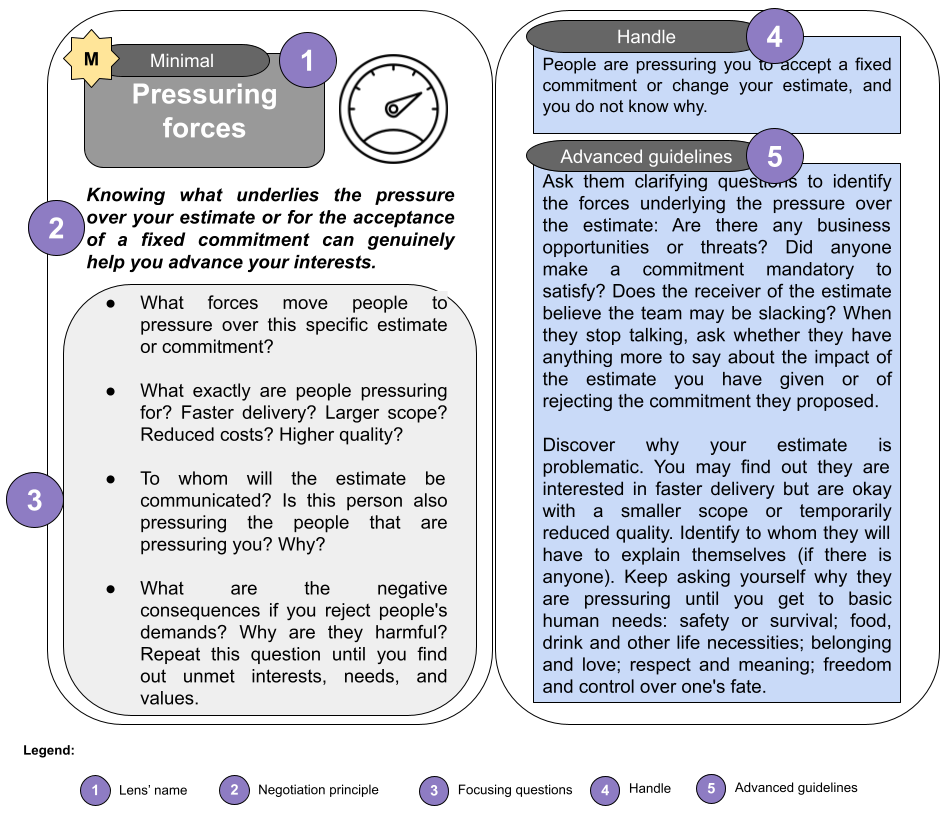}
  \caption{Picture of one of the lenses. A larger and richer version is available at\url{https://doi.org/10.6084/m9.figshare.16815115.v2}.}
  \Description{Picture of one of the lenses. A larger version is available at \url{https://doi.org/10.6084/m9.figshare.16815115.v2}.}
  \label{fig:lenses}
\end{figure}

To satisfy the objective of empowering software practitioners in the defense of their software estimates and the negotiation of commitments, we adopted the Design Science Research (DSR) approach \cite{wieringa_design_2014}, as we describe in \cite{fernandes_matsubara_dealing_2019}. We designed an artifact: a set of defense lenses covering the principles and steps from all the negotiation methods that form our foundations (Section \ref{sec:back}). We described the whole set of lenses and their supporting examples in a booklet, which we are now improving as part of our design cycle in DSR. In this section, we present only one lens from our set --- presented in Figure \ref{fig:lenses} --- to illustrate the artifact.

The lenses' format was inspired by the design lenses proposed to guide the design of gamified systems \cite{deterding_lens_2015}. Design lenses combine a name, a design principle, and a set of focusing questions, supporting the designer to take a mental perspective regarding the design issue the lens focus on. Likewise, the defense lenses also have a name (Figure \ref{fig:lenses}.1), a negotiation principle on which we based them (Figure \ref{fig:lenses}.2), and a set of focusing questions (Figure \ref{fig:lenses}.3) on the front side. The focusing questions aim to provoke a mindset shift in estimators: from changing estimates to defending them if no legitimate reasons justify a change, and negotiating commitments that others are trying to impose when they are unrealistic, instead of accepting them promptly. On the backside, we enriched the lens with a handle (Figure \ref{fig:lenses}.4), describing the situations to apply that lens. It helps identify the specific lens we need when using them in isolation. We also added advanced guidelines (Figure \ref{fig:lenses}.5) to support less experienced practitioners or help in more complex situations. 

The lens that we exemplified in Figure \ref{fig:lenses} is entitled \textit{Pressuring Forces}. It is based on the negotiation principle of \textbf{focus on interests, not positions} \cite{fisher_getting_2011} and of listening attentively to the other side to understand their perspective fully –-- as part of the \textbf{respect your way to Yes} step \cite{ury_power_2012}. The first focusing question aims to investigate the pressuring side's needs and interests, while the second focuses on understanding precisely the what they are pressuring for. The third focusing question regards a particular pressuring force: when someone more powerful is pressing the other side. Identifying this can help the estimator to \textbf{step to their side} \cite{ury_getting_2007}, by opening eyes to ways to relieve their pressure. The last focusing question aims to make the estimator get perspective on the negative consequences that can emerge if the other side’s interests are not satisfied, relating it to the \textbf{respect your way to Yes} step \cite{ury_power_2012}. The advanced guidelines further explore what and how the estimator can investigate the other side's interests. 

Except for the Wildcard, which is a particular lens, we organized all the others in two packs: (i) the Minimal Defense Pack and (ii) the Extended Pack. The first encompasses five lenses with core guidance for dealing with specific pressure episodes. The latter provides additional guidance for the situations when the estimator already tried to deflect from a pressure episode using the the Minimal Defense Pack. However, people keep pushing for unjustified changes in the estimates or the acceptance of an unattainable commitment.

Our artifact is also composed of examples that can help even further estimators. For the Pressuring forces lens, we provided a table with a set of detailed examples of clarifying questions that can be useful, including justifications and additional follow-up questions. For instance, a clarifying question is "Does it look like our team can be more productive than this estimate suggests?". The justification is "People may believe the team is unnecessarily padding the estimates, or that lower estimates produce healthy pressure, making the team more productive". The follow-up advice is "Break the tasks to clarify the work dimension". 

\section{Preliminary Results}

\begin{table*}[htb!]
  \caption{Participants' profile and lenses' uses. The median value from participants is in italics. Values above median are underlined. U = Lens the participant used. Nu = Lens the participant did not understand. Na = Lens the participant decided not to apply. OpE = openness to experience. Con = conscientiousness. Ext = extroversion. Agr = agreeableness. Neu = neuroticism. AsL = assertiveness level. PId = Participant's Id. Y = Years working in the software industry with software engineering activities. Wild = Wildcard. Ladd = Laddering whys. Pres = Pressuring forces. Choo = Choose your battles. Asser = Assert your estimates. Cand = Candidate commitments. Keep = Keep strategy. Pers = Perspective taking. Real = Reality test. Gold = Golden Bridge.}
  \label{tab:profile}
  \begin{tabular}{|p{0.6cm}|p{0.6cm}|p{0.5cm}|p{0.5cm}|p{0.5cm}|p{0.5cm}|p{0.5cm}|p{0.3cm}|p{0.7cm}|p{0.7cm}|p{0.7cm}|p{0.7cm}|p{0.7cm}|p{0.7cm}|p{0.7cm}|p{0.7cm}|p{0.7cm}|p{0.7cm}|}
    \toprule
    \textbf{OpE} & \textbf{Con} & \textbf{Ext} & \textbf{Agr} & \textbf{Neu} & \textbf{AsL} & \textbf{PId} & \textbf{Y} & \textbf{Wild} & \textbf{Ladd} & \textbf{Pres} & \textbf{Choo} & \textbf{Asser} & \textbf{Cand} & \textbf{Keep} & \textbf{Pers} & \textbf{Real} & \textbf{Gold} \\
    \midrule
     \textit{15} & \textit{17} & 11 & \textit{16} & \underline{15} & 43 & \textbf{P1} & 23  & Nu & U & U & U & U & U & Na &  &  & Nu \\
     14 & 14 & \textit{15} & 14 & 12 & \textit{51} & \textbf{P2} & 1 & Nu &  & Nu &  & U & U &  &  &  &  \\
     6 & 11 & 8 & 14 & \underline{17} & 42 & \textbf{P3} & 1 &  &  &  & U & U & U &  &  &  &  \\
     \underline{20} & \underline{19} & \underline{17} & \underline{20} & \textit{13} & \underline{69} & \textbf{P4} & 12 & Nu & U & Na & Na & U & U & Na &  & Na & \\
     \cline{9-18}
     \multicolumn{1}{|p{0.6cm}}{\underline{17}} & \multicolumn{1}{|p{0.6cm}}{\underline{18}} & \multicolumn{1}{|p{0.5cm}}{\underline{19}} & \multicolumn{1}{|p{0.5cm}}{\underline{18}} & \multicolumn{1}{|p{0.5cm}}{11} & \multicolumn{1}{|p{0.5cm}}{\underline{57}} & \multicolumn{1}{|p{0.5cm}}{\textbf{P5}*} & \multicolumn{1}{|p{0.3cm}}{23} & \multicolumn{10}{|c|}{*P5 acted as the receptor of estimates.} \\
    \hline
    
  \end{tabular}
\end{table*}

In this section, we present the preliminary results regarding the evaluation of our artifact. We carried out a focus group \cite{kontio_focus_2008}, focused on understating about (i) the \textbf{perceived usefulness} and (ii) \textbf{improvement opportunities} for the design lenses. We chose the focus group before using other evaluation methods, because it would enable us to improve the lenses at a lower cost/risk for software professionals. We selected participants from our network, focusing on covering different perspectives in terms of experience and roles. Thus, we invited people with varied experience with software engineering in terms of years of experience (ranging from one to 23 years) and roles (including people with experience as software developers, testers, and managers). 

In the first focus group meeting, we described the study and the artifact to participants, also providing the booklet for study. In the second meeting, we started by resolving questions from participants and collecting their improvements suggestions. Next, participants engaged in a role-playing activity guided by one scenario similar to the one described in \cite[p.~5]{mcconnell_what_2006}\footnote{Supplementary material with the focus group script (including the scenario) is available at \url{https://doi.org/10.6084/m9.figshare.16815115.v2}.}. One participant played the receptor role (P5 in Table \ref{tab:profile}), and the others acted as estimators. The estimators worked together to estimate a software project entitled SeminarWeb \cite{jorgensen_identification_2010}, which we chose after searching for detailed specifications used in previous studies about software effort estimation. Next, all participants got together to define a commitment about the software project, and the receptor pressured estimators to accept a commitment that their estimate did not support, adhering to the scenario description. For this scenario, we expected participants to use the lenses from the Minimal Pack, specially the \textit{Assert your Estimate}, \textit{Pressuring Forces}, and \textit{Candidate Commitments} lenses. After the participants decided they reached the end of the commitment establishment phase, they answered a questionnaire about the lenses and we had a debriefing session for more improvement suggestions.

We had five participants from the software industry, represented by a participants' ID in Table \ref{tab:profile}. We also collected data regarding participants' \textbf{personality traits} (Big-Five Inventory, with 20-items \cite{veloso_gouveia_short_2021}) and \textbf{assertiveness} levels (Rathus Assertiveness Scale \cite{rathus_30-item_1973}), because these can be intervening variables in our study. We present participants' profiles and their lenses' uses in Table \ref{tab:profile}.

Table \ref{tab:profile} shows that participants decided to use all lenses from the Minimal Pack, specially the \textit{Assert your estimate} and the \textit{Candidate commitments}, as we expected. This indicates that they perceived the set of lenses as useful given the scenario at hand. Unexpectedly, only P1 chose to use the \textit{Pressuring Forces} lens, and P4 thought it would be better not to apply it at all (Na in Table \ref{tab:profile}). Also, P2 did not understand it (Nu in Table \ref{tab:profile}), revealing a need to improve this lens. Also, as we expected, participants decided not use the lenses from the Extended Pack. P4 informed explicitly that he did not apply some of the lenses of this pack, because he believed they would be useful only for inflexible scenarios when more steady guidance is needed---showing that he understood the purpose of this pack well. 

A few participants indicated they did not understand the \textit{Wildcard}. P1 explained that she had difficulty getting the idea because the wildcard metaphor was broken: when learning a game, one usually learn all its rules and details. The wildcard is presented at the end as a versatile card. However, in our booklet, it was the first card to be presented and served as a guide for some of the other ones. Participants also indicated other improvement opportunities for the presentation format: the current booklet format is helpful for training purposes, but it lacks an indexing system to quickly search for the right lens when the situation requires the estimators to respond fast. We designed the handle and the Wildcard to provide such aid. Nevertheless, these elements were not entirely effective. These preliminary results showed that participants perceived the defense lenses as useful and to a large extent understandable, although improvements are still needed.

Another interesting result comes from P3---the estimator with low assertiveness, and one of the least experienced participants. He reported using few cards, possibly explained by the low score in the openness trait. In addition, he discussed during the debriefing that he does not feel comfortable arguing with other people, specially more experienced ones. He believes that example sentences that the estimator can pick from the lens and use directly could help overcome this. Interestingly, P4---the estimator with a high assertiveness profile---made a counterpoint that the set of lenses as a whole helps the estimator to gather arguments together. This shows that even people who already have an appropriate assertive behavior can benefit from the lenses in the estimation context. 

\section{Future Plans}

In this paper we discuss the relevance of tackling the problem that software practitioners deliberately change their estimates, yielding to pressure over their estimates during the establishment of commitments instead of defending them. We proposed a set of defense lenses: a supporting artifact to empower estimators to deal with this problem. The preliminary results from a focus group provide evidence that the proposed defense lenses are perceived as useful and usable, even for estimators with an assertive profile. Moreover, this endeavor is innovative because it is a concrete step towards supporting software practitioners in developing the relevant soft skill of negotiation, which can be useful in software engineering activities beyond software estimation.

Our next step is to redesign the lenses to include the identified improvements, and to conduct other focus groups, to look for more improvement opportunities. Then, we plan to develop a digital simulation-based training to aid practitioners in acquiring the negotiation skills through the lenses. Digital simulations are promising in promoting self-efficacy beliefs and transfer of training to work \cite{gegenfurtner_digital_2014}. Next, we plan to carry out a field experiment \cite{stol_guidelines_2020}, to collect empirical evidence on the practitioners' perceived usefulness of adapting the negotiation methods in the form of the defense lenses in the field, to reach a conclusion about whether our artifact can actually empower them in the defense of their software estimates.

\begin{acks}
This research, according to Article 48 of Decree nº 6.008/2006, was funded by Samsung Electronics of Amazonia Ltda, under the terms of Federal Law nº 8.387/1991, through agreement nº 003, signed with ICOMP/UFAM. Also supported by CAPES - Financing Code 001, CNPq processes 314174/2020-6 and 313067/2020-1, FAPEAM process 062.00150/2020, and grant \#2020/05191-2 S\~{a}o Paulo Research Foundation (FAPESP).
\end{acks}

\bibliographystyle{ACM-Reference-Format}
\bibliography{nier-arxiv}

\end{document}